\documentclass[12pt,a4paper]{article}

\usepackage[british]{babel}

\usepackage[a4paper,top=2cm,bottom=2cm,left=2.5cm,right=2.5cm,marginparwidth=1.75cm]{geometry}

\usepackage[numbers]{natbib}
\usepackage{amsmath}
\usepackage{graphicx}

\usepackage{orcidlink}
\usepackage[title]{appendix}
\usepackage{mathrsfs}
\usepackage{amsfonts}
\usepackage{booktabs} % For \toprule, \midrule, \botrule
\usepackage{caption}  % For \caption
\usepackage{algorithm}
\usepackage{listings}
\usepackage{enumitem}
\usepackage{chngcntr}
\usepackage{lipsum}
\usepackage{subcaption}
\usepackage{authblk}
\usepackage{bm}
\usepackage[T1]{fontenc}    % Font encoding
\usepackage{csquotes}       % Include csquotes
\usepackage{diagbox}
\usepackage{hyperref}

\usepackage{setspace}
\usepackage{titlesec}
\usepackage{float}   % for customizing caption position
\usepackage{caption} % for customizing caption format



\title{Thin-shell wormhole with a background Kalb-Ramond Field}

\author[1]{Arya Dutta}
\author[2]{Farook Rahaman}

\affil[1]{\small  Yeshiva University, Department of Mathematical Sciences, New York, NY 10016, USA\newline
Email: adutta1@mail.yu.edu}
\affil[2]{ Jadavpur University, Department of Mathematics, Kolkata - 700032, India \newline Email: farook.rahaman@jadavpuruniversity.in }

\date{}  % Remove date

\begin{document}
\maketitle 

\begin{abstract}
The Kalb–Ramond field is a background tensor field that arises in string theory and violates the local Lorentz symmetry of spacetime, upon acquiring the Vacuum Expectation Value (VEV). A non-minimal coupling between the Kalb–Ramond VEV and the Ricci tensor may give rise to a modified black hole solution. Considering two copies of such black holes, we construct a thin-shell wormhole using the `Cut-and-Paste' technique. Then we investigate key physical properties of the wormhole like pressure-density profile, equation of state, the geodesic motion of test particles near the wormhole throat, and the total amount of exotic matter in the throat, and examine how these properties vary with the Lorentz-Violating (LV) parameters. We find that the wormhole model violates the null and weak energy conditions, but satisfies the strong energy condition. On top of that, the velocity of the throat radius is found considering its time evolution. Finally, we analyze its linear stability against small radial perturbations. 
\end{abstract}


%-------------------------------------------
% Paper Body
%-------------------------------------------
%--- Section ---%
\section{Introduction}\label{Intro}
A wormhole is physically a hypothetical shortcut between two distant regions in spacetime, and mathematically a solution of the Einstein equations. The term `wormhole' was first used by John Archibald
Wheeler and C.W. Misner in their paper \cite{MISNER1957525} in 1957. In reality, the history of developing the concept of wormholes goes back to 1935, when Albert Einstein and Nathan Rosen \cite{1} tried to formulate a potential atomistic theory of matter and electricity, and proposed a wormhole-like solution called the `Einstein-Rosen bridge'. However, J.A. Wheeler and Robert W. Fuller \cite{2} argued that such a bridge, if it connects two parts of the same universe, is too unstable to let a signal pass through it. It was not until 1988 that a stable `two-way' traversable wormhole was found as a solution to Einstein's field equations. Proposed by Michael S. Morris and Kip S. Thorne \cite{3}, this wormhole is a hypothetical tunnel between two regions of the same universe or different universes, and its throat is a hypersurface of minimal area. Additionally, they showed that holding the wormhole open requires exotic matter, which violates the Null Energy Condition (NEC) $T_\mu{}_\nu k^\mu k^\nu\ge0$, $k_\mu k^\nu=0$. So, one should aim to minimize the usage of exotic matter while theoretically constructing a wormhole. One way to make it infinitesimally small is to choose a suitable geometry \cite{4}. Since the past few decades, traversable wormholes have been thoroughly explored in modified gravity theories like torsion-like Rastall gravity \cite{WAHEED2025170078}, $f(R, T)$ gravity \cite{doi:10.1142/S0219887825501877, sharif2023traversablewormholesolutionsadmitting, KiroriwalKumarMaurya2024}, noncommutative geometry \cite{Kuhfittig_2013, PhysRevD.86.106010}, and so on.

Matt Visser \cite{5} came up with another interesting way of reducing the amount of exotic matter by concentrating it in the wormhole's throat. In this method, known as the `Cut and Paste' technique or the surgical technique, two portions from two different blackhole spacetimes are cut and pasted in such a way that no event horizon can be formed. Thus, the newly obtained manifold is geodesically complete, where two asymptotically flat spacetimes are joined by a three-dimensional timelike hypersurface, which is called the `thin shell' or the throat. This new construction of a wormhole, known as the thin-shell wormhole (TSW), is extremely useful for performing dynamic stability analysis. The main difference between a TSW  and a traversable wormhole is that the former holds the exotic matter only at the thin shell, while it is distributed in a finite region around the throat in the latter. The surface stress-energy tensor of the exotic matter at the thin shell and the geometry of the throat can be analyzed, respectively, using Darmois-Israel formalism \cite{6,7}, and junction condition formalism \cite{8,9}. Poisson and Visser \cite{10} have analyzed the linearization stability of a thin wormhole constructed by joining
two Schwarzschild spacetimes.

Over the past few decades, many authors have adopted this surgical technique to construct thin-shell wormholes for different settings. For example, TSWs have been analyzed in several modified gravities, like dilaton gravity \cite{PhysRevLD.71.127501}, Einstein–Maxwell theory with a Gauss–Bonnet term \cite{TSW_EMT_GB}, Hořava–Lifshitz gravity \cite{18}, Einstein-Gauss-Bonnet gravity \cite{EGB_grav}, Brans–Dicke gravity \cite{TSW_Brans–Dicke_gravity}, dilaton-axion gravity \cite{usmani2010thin}, Heterotic string theory \cite{Heterotic_string_th}, brane-world gravity \cite{TIdalCharged_BH}, teleparallel gravity \cite{JavedMustafa2022}, scalar fields \cite{scalar_field_TSW_BWG}, $f(R)$ gravity \cite{GODANI2022101835, Figueroa_Aguirre_2023}; and also non-commutative geometry \cite{Sharif_2012, Kuhfittig_2012, Bhar_2015}. Since the usual presence of exotic matter tends to make a  TSW unstable, their linear, non-linear, and thermodynamic stability has also been a topic of great interest for scientists \cite{Eiroa_2024, PhysRevD.76.24021, PhysRevD.71.124022, Mazharimousavi_2024}. It has been observed that excessive violation of the NEC can cause instabilities of a wormhole under radial perturbation, but suitable choices of equations of state, such as in Chaplygin gas TSWs \cite{PhysRevD.76.24021}, TSWs with polytropic matter \cite{Sharif_2012, 2023NewA..10102021E, osti_21409723, sharif2016stabilityregularhaywardthinshell} can help to construct stable TSW as well.

There is another interesting astrophysical object called Gravastar, which shares a similar construction to TSW, and is worth discussing here. Gravastars (gravitational vacuum stars), first proposed by Mazur and Mottola \cite{mazur2001gravitational}, are compact objects in which an interior de Sitter spacetime is matched to an exterior Schwarzschild geometry through a thin shell of matter. They are considered as alternatives to black holes, avoiding singularities and horizons. Both Gravastar and TSW rely on the Israel junction formalism \cite{6, 7, 8} to glue distinct spacetimes across a matter shell, but exotic matter is not required for a gravastar to form. Recent developments in the research on gravastars include exploring them in modified gravities like $f(Q)$ gravity \cite{Mohanty_2024}, $f(R, T)$ gravity \cite{SALEEM2024169659}, massive gravity \cite{DAS2024101691}, and noncommutative geometry \cite{Silva_2024}.

On the other hand, Lorentz Symmetry Breaking (LSB) has been quite a famous idea in the contexts of string theory \cite{Spontaneous_LSB, PhysRevL.63.224227, PhysRevL.66.18111814}, loop quantum gravity \cite{PhysRevL.84.23182321, PhysRevD.65.103509}, various other modified theories of spacetime symmetry \cite{PhysRevL.97.021601, Dirac_equation_STR, PhysRevL.88.190403}, and modified gravity theories \cite{NCFT_LV, PhysRevD.79.084008}. The local LSB can be driven by a spontaneous symmetry-breaking potential due to self-interacting tensor fields \cite{PhysRevD.69.105009}. The simplest tensor field of this kind is a vector field and is known as the bumblebee field \cite{PhysRevD.69.105009, PhysRevD.71.065008}. In this bumblebee gravity, black hole solutions have already been found \cite{PhysRevD.97.104001, PhysRevD.72.044001}, followed by deriving traversable wormhole solutions as well \cite{PhysRevD.99.024042, oliveira2019quasinormal}. 

The Kalb–Ramond (KR) field,  denoted by $B_\mu{}_\nu$, and proposed by M. Kalb and P. Ramond in 1974 \cite{45}, is another self-interacting antisymmetric 2-tensor, which arises in the bosonic string theory. This field can violate Lorentz Symmetry when the Vacuum Expectation Value (VEV) is achieved. Considering a Lorentz invariance, a hairy black hole solution \cite{PhysRevD.53.2244} was found for minimal coupling between the KR field and gravity, where the event horizon is turned into a naked singularity. Recently, L. Lessa, J. Silva, R. Maluf, and C. Almeida1 derived an exact spherically symmetric and static, modified black hole solution \cite{Lessa_2020}, considering non-minimal coupling between the spacelike Kalb-Ramond Field and the Ricci tensor. Soon after, L.Lessa and their collaborators also found a traversable wormhole solution in this field \cite{LESSA2021168604}. What is very interesting in the modified black hole solution in \cite{Lessa_2020} is that for a particular choice of the LSB parameter, the Lorentz violation produces a solution similar to the Reissner–Nordstrom, despite the absence of charge. 

In the present paper, we extend their work to formulate a thin-shell wormhole in this background tensor field, called the Kalb-Ramond (KR) field, based on the works cited above. We consider two copies of the modified, hairy black holes in the KR field and construct a thin-shell wormhole by the `Cut and Paste technique' described earlier \cite{5}. Then we go on to investigate some of the most important properties of this newly constructed wormhole, both analytically and graphically, and also examine how these phenomena are affected by the change of the Lorentz-Violating (LV) parameters $\lambda$ and $\gamma$. This analysis gives us valuable insights about the wormhole, like checking for energy conditions, finding the equation of state, concluding about the attractive or repulsive nature of its gravitational field, the distribution of exotic matter in the structure, and its linear stability against small radial perturbations. We perform this stability analysis in terms of the speed of sound parameter, $\beta$. We find that the wormhole model is mostly unstable under the assumption that $0<\beta^2\leq 1$, which is true for regular matter. However, this is uncertain since exotic matter is involved, and $\beta$ might take other values for exotic matter, which we do not know yet.

The paper is organized as follows: We look at the features of the black hole in the KR field in Section \ref{sec: BH sol}, and thus prepare the groundwork for constructing our wormhole model and investigating the components of stress-energy around its throat in Section \ref{sec: TSW constr.}. The Equation of State (EOS) at the thin-shell and the motion of a test particle near the shell are studied, respectively, in Section \ref{sec: omega} and Section \ref{a^r}. In Section \ref{Sec: Rad. Vel}, we look at the nature of the time evolution of the radius of the throat. Calculating the total exotic matter content at the throat in Section \ref{Omega}, we analyze our model's linearized stability in Section \ref{beta}. Finally,
in Section \ref{Concl.} we provide some concluding remarks.

\section{Modified Black Hole Solution in The Kalb-Ramond Field}\label{sec: BH sol}
Assuming a spacelike Kalb–Ramond field with a constant squared norm VEV and its non-minimal coupling with the
Ricci tensor, the following modified black hole solution was found in  \cite{Lessa_2020} :
\begin{equation}
    ds^2= -f(r)dt^2 + f(r)^{-1}dr^2 + r^2d\theta^2 +r^2sin^2\theta d\phi^2 ,
\end{equation}
where
\begin{equation}
    f(r)= 1 - \frac{R_s}{r} + \frac{\gamma}{r^{(2/\lambda)}}.
\end{equation}
\begin{figure}[htbp]
\centering
      \begin{subfigure}[b]{0.32\textwidth}
       
        \includegraphics[width=\textwidth]{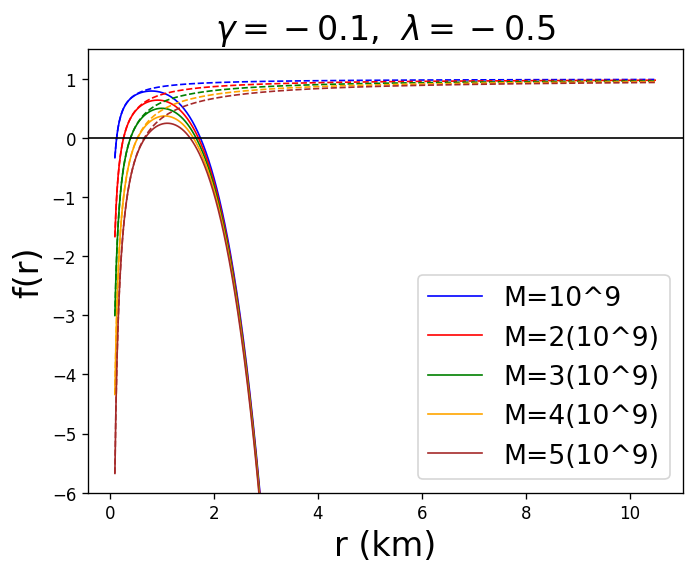}

      \caption{} 
     \end{subfigure}
    \hfill
    \begin{subfigure}[b]{0.32\textwidth}
    
        \includegraphics[width=\textwidth]{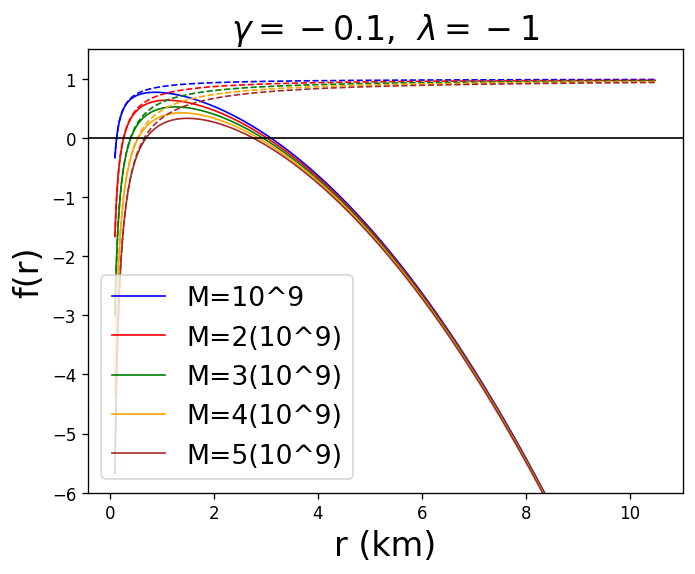}
    \caption{}
    \end{subfigure}
    \hfill
     \begin{subfigure}[b]{0.32\textwidth}
       
        \includegraphics[width=\textwidth]{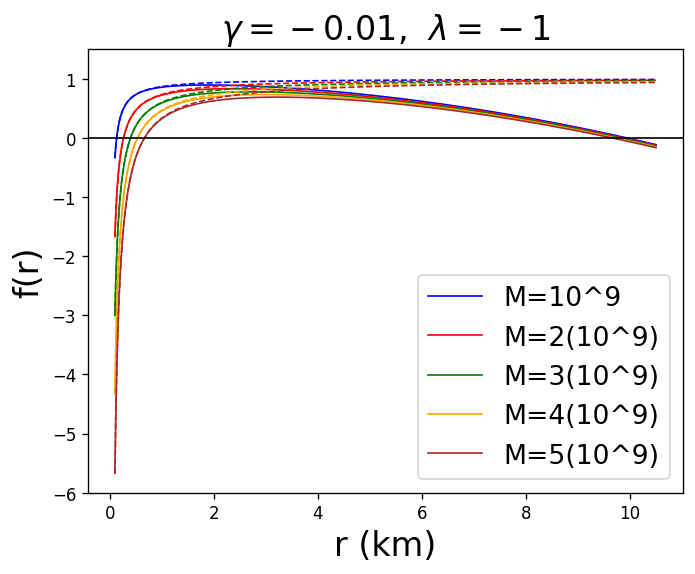}
       \caption{}
    \end{subfigure}
    \caption{The variation in the metric function $f(r$) with the radial coordinate $r$ is shown. We have taken three sets of values for the LV parameters $\gamma$ and $\lambda$ (which meet our given constraints) and plotted $f(r)$ (solid lines) for five different masses (M varies from $10^9$ to $5.10^9$ Kg) in each of these three cases. The dashed lines represent Schwarzschild Black Holes corresponding to each M, $\gamma$, and $\lambda$. Note that the r-intercepts of $f(r)$ (the points where $f(r)$=0) for each curve are the radial distances at which event horizons form. Since each curve cuts the r-axis at two points, each black hole has two event horizons (inner and outer, occurring at $r_-$ and $r_+$ respectively).} \label{fig:f(r)}
\end{figure}

Here $R_s$ = $2GM$ = the usual Schwarzschild radius (where $M$ = the mass of the corresponding black hole), and $\gamma$ and $\lambda$ are Lorentz-Violating (LV) parameters. To be more specific,
$\lambda$:= $|b|^2\xi_2$ (where $|b|^2$ is the constant norm of the Kalb-Ramond VEV $b_\mu{}_\nu$, and $\xi_2$ is a non-coupling constant), and
$\gamma$ is a constant (with dimensions [$\gamma$] = $L^{2/\lambda}$)  that controls the Lorentz violation effects upon the Schwarzschild solution. We can see that if $\gamma$=0, then $f(r)$ represents nothing but a Schwarzschild Black Hole.

Analyzing the equation of state and energy conditions of the Lorentz-violating source, the authors of  \cite{Lessa_2020} concluded that the Weak Energy Condition is satisfied for $0 \leq \lambda \leq 2 $ and $\gamma > 0$. Now, we know that the wormhole's thin shell (throat) comprises exotic matter, and exotic matter violates the Weak Energy Condition and the Null Energy Condition (NEC). Hence, in the case of studying properties (pressure, energy density, gravitational field, stability, etc.) of the throat, we will consider the values of the LV parameters out of the above range, i.e., $\lambda \in (-\infty, 0) \bigcup  (2, \infty)$ and $\gamma \leq 0 $.

From \autoref{fig:f(r)}, we observe that for a decrease in $\lambda$ (compare (a) and (b)) and an increase in $\gamma$ (compare (b) and (c)), the extremal distance (the distance between the inner and outer event horizons) increases. Moreover, for a particular set of the LV parameters, the smaller the mass of the black hole, the larger the extremal distance. The plot also indicates that after a certain value of r, the metric function f(r) is not mass-dependent anymore. Finally, we note that each of these hairy black hole solutions has two horizons, except the Schwarzschild cases (non-hairy), which have only one.

\section{Thin-Shell Wormhole Construction}\label{sec: TSW constr.}
A thin-shell wormhole can be formed by joining two distinct black hole spacetimes through a cut-and-paste technique, independent of the underlying gravitational theory. This process employs the Israel–Darmois junction conditions to ensure a smooth matching at the interface. 
The geometric formalism, grounded in General Relativity, remains valid and maintains the continuity of the spacetime metric. Consequently, this approach provides a versatile and robust framework for constructing 
thin-shell wormholes, even within the context of alternative or modified theories of gravity.

We consider two copies of the black hole described in Section \ref{sec: BH sol} and remove the following four-dimensional regions from each:
\begin{equation}
    \Omega^{\pm} \equiv \{r \leq a \,|\, a > r_h\},
\end{equation}
where a is a constant, and $r_h$ refers to the event horizon, here $r_h$ = $r_+$, i.e. the larger of the two radii.
After this `cutting', we are left with two asymptotically flat regions and two timelike hypersurfaces. These hypersurfaces can be topologically represented as: 
\begin{equation}
    \delta\Omega^{\pm}= \{r = a \,|\, a > r_h\}.
\end{equation}
Now, we glue them and form the throat (or the thin-shell), denoted by $\Sigma$, at  $\delta\Omega^+$= $\delta\Omega^-$= $\delta\Omega$, connecting the asymptotically flat spacetimes. This operation completes our construction of a new, geodesically complete manifold, termed the thin-shell wormhole. The induced metric on $\Sigma$ is given by
\begin{equation}
    ds^2= -d\tau^2 + a^2(\tau)d\Omega_D{}^2,
\end{equation}
where $\tau$ is the proper time at the junction shell and a($\tau$) is the radius of the throat. Using Lanczos
equation \cite {
PhysRevLD.71.127501, PhysRevD.70.044008, lanczos1924flachenhafte,perry1992traversible,sen1924grenzbedingungen}, we find the intrinsic surface stress-energy tensor as
\begin{equation}
    S^i_j = -\frac{1}{8\pi} ([K^i_j] - \delta^i_j[K]).
\end{equation}
$S^i_j$ can be obtained as diag($-\sigma$, $p_\theta$, $p_\phi$), where $\sigma$ is the surface-energy density and
$p = p_\theta = p_\phi$ is the surface pressure. Again applying the Lanczos equations,
\begin{equation}
    \sigma = -\frac{1}{4\pi}[K^\theta_\theta]
\end{equation}
and
\begin{equation}
    p = \frac{1}{8\pi}([K^\tau_\tau] + [K^\theta_\theta]).
\end{equation}
Now, assuming that the radius $r=a$ is a function of time, the following dynamic analysis \cite{10} is found 

\begin{equation}
    \sigma = -\frac{1}{2\pi a}\sqrt{ f(a)+\dot{a}^2}
\end{equation}\label{eq: sigma}
and
\begin{equation}
    p = -\frac{\sigma}{2} + \frac{2\ddot{a}+f'(a)}{8\pi\sqrt{ f(a)+\dot{a}^2}},
\end{equation}\label{eq: p}
where the overdot and prime denote, respectively, the derivatives with respect to $\tau$ and $a$. Here $p$ and $\sigma$ obey the energy conservation equation
\begin{equation}
    \frac{d}{d\tau}(\sigma a^2) + p\frac{d}{d\tau}(a^2)=0,
\end{equation}
or,
\begin{equation}
    \dot{\sigma}+2(\sigma+p)\frac{\dot{a}}{a}=0.
\end{equation}\label{eq 12}

For a static configuration of radius $a$ (i.e. $\dot{a}$ = 0 and $\ddot{a}$ = 0), we obtain from Eqs. \ref{eq: sigma} and \ref{eq: p},
\begin{equation}
     \sigma=-\frac{1}{2\pi a} \sqrt{1-\frac{2GM}{a} + \frac{\gamma}{a^{2/\lambda}}}
\end{equation}\label{eq: sigma 1}
and
\begin{equation}
    p=\frac{1}{4\pi a} \frac{1-\frac{GM}{a} + (1- \frac{1}{\lambda})\frac{\gamma}{a^{2/\lambda}}}{\sqrt{1-\frac{2GM}{a} + \frac{\gamma}{a^{2/\lambda}}}}.
\end{equation}\label{eq: p1}

\begin{figure}[htbp]
\centering
      \begin{subfigure}[b]{0.32\textwidth}
       
        \includegraphics[width=\textwidth]{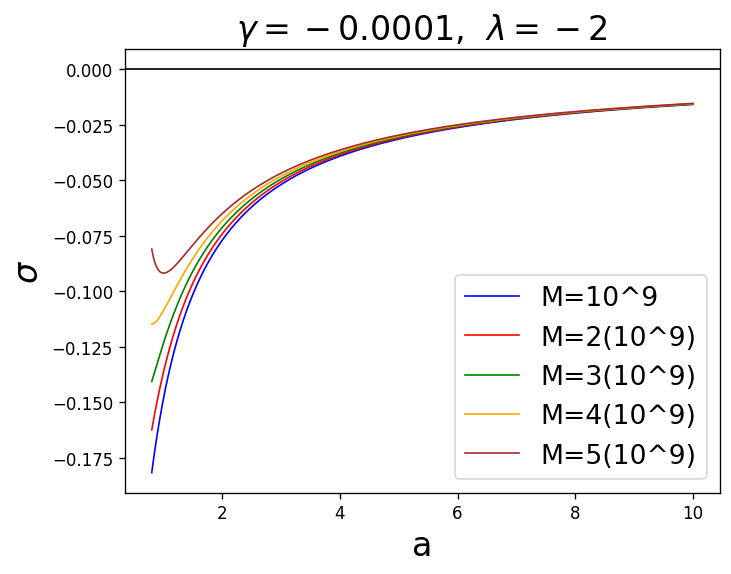}

      \caption{} 
     \end{subfigure}
    \hfill
    \begin{subfigure}[b]{0.32\textwidth}
    
        \includegraphics[width=\textwidth]{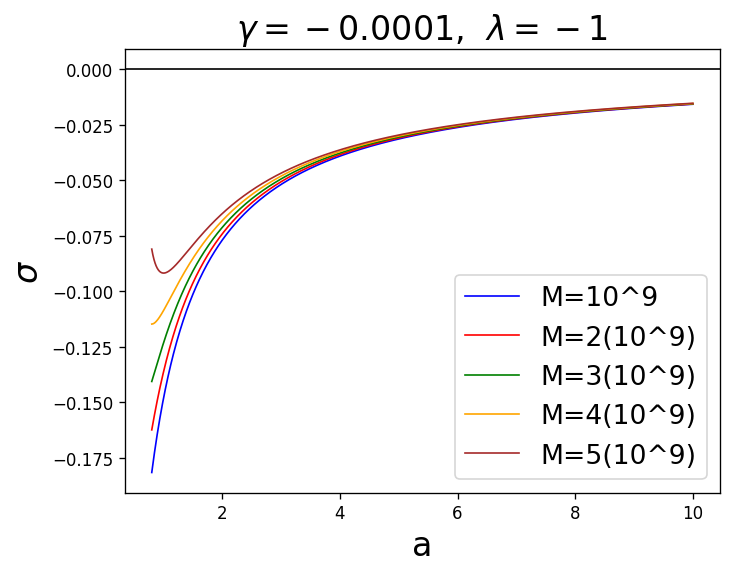}
    \caption{}
    \end{subfigure}
    \hfill
     \begin{subfigure}[b]{0.32\textwidth}
       
        \includegraphics[width=\textwidth]{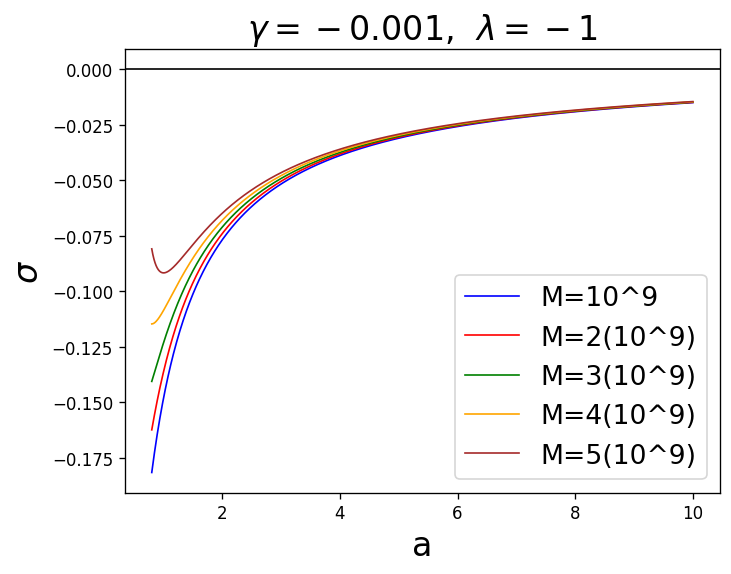}
       \caption{}
    \end{subfigure}
    \caption{The variation of the energy density \textbf{$\bm{\sigma}$} with the throat radius \textbf{$\bm{a(km)}$}, for different masses and LV parameters. We choose typical wormholes whose radii fall within the range of 1 to 10 km. Since $\sigma<0$ from these plots, the first condition for the Weak Energy Condition (WEC) is violated.} 
    \label{fig:sigma}
\end{figure}

\begin{figure}[htbp]
\centering
      \begin{subfigure}[b]{0.32\textwidth}
       
        \includegraphics[width=\textwidth]{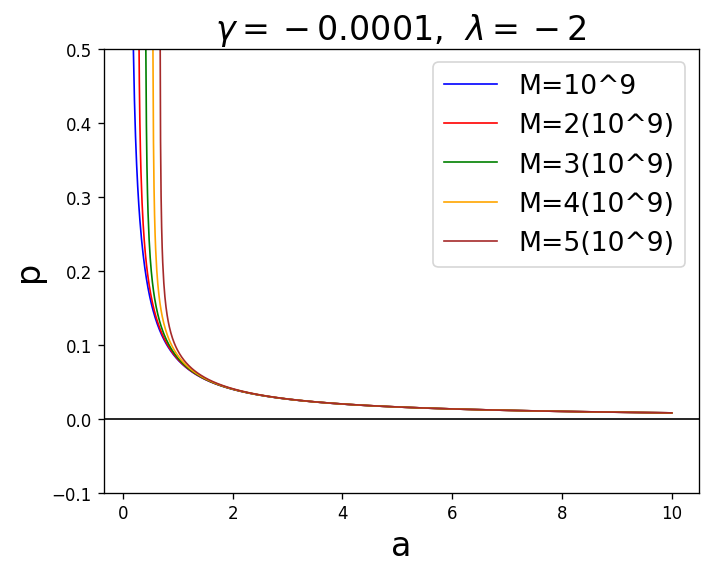}

      \caption{} 
     \end{subfigure}
    \hfill
    \begin{subfigure}[b]{0.32\textwidth}
    
        \includegraphics[width=\textwidth]{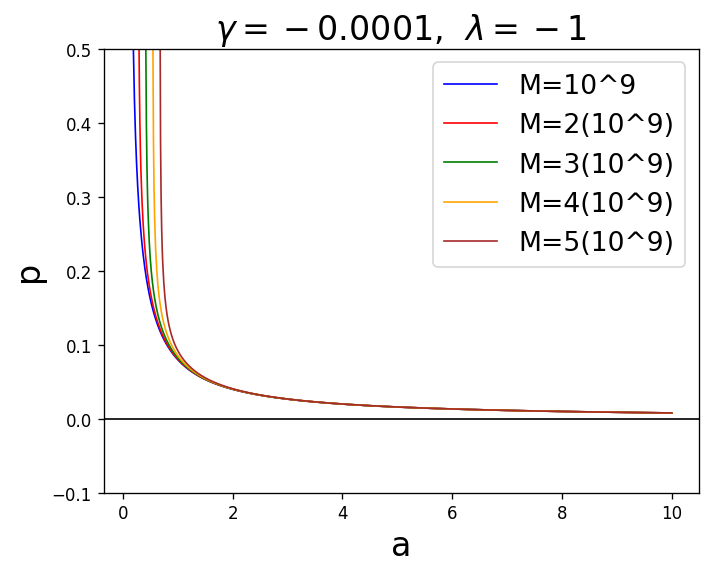}
    \caption{}
    \end{subfigure}
    \hfill
     \begin{subfigure}[b]{0.32\textwidth}
       
        \includegraphics[width=\textwidth]{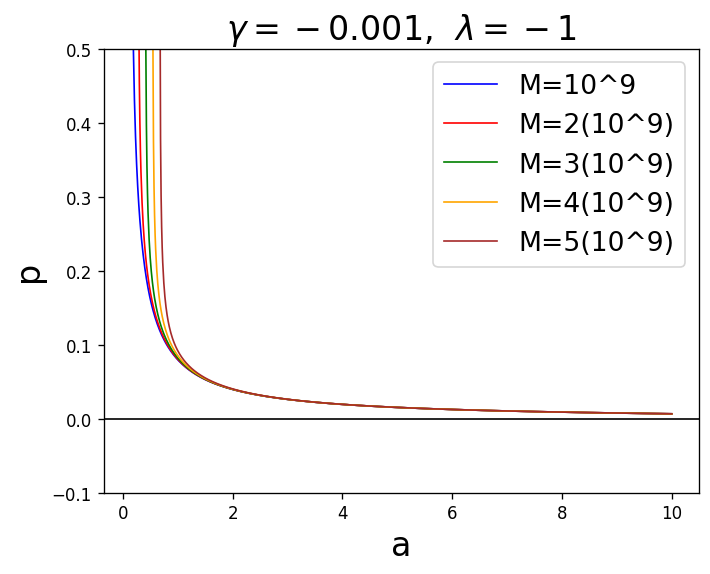}
       \caption{}
    \end{subfigure}
    \caption{The variation of the thermodynamic pressure \textbf{$\bm{p}$} with the throat radius \textbf{$\bm{a(km)}$}, for different masses and LV parameters.} 
    \label{fig: p}
\end{figure}

\begin{figure}[H]
\centering
      \begin{subfigure}[b]{0.32\textwidth}
       
        \includegraphics[width=\textwidth]{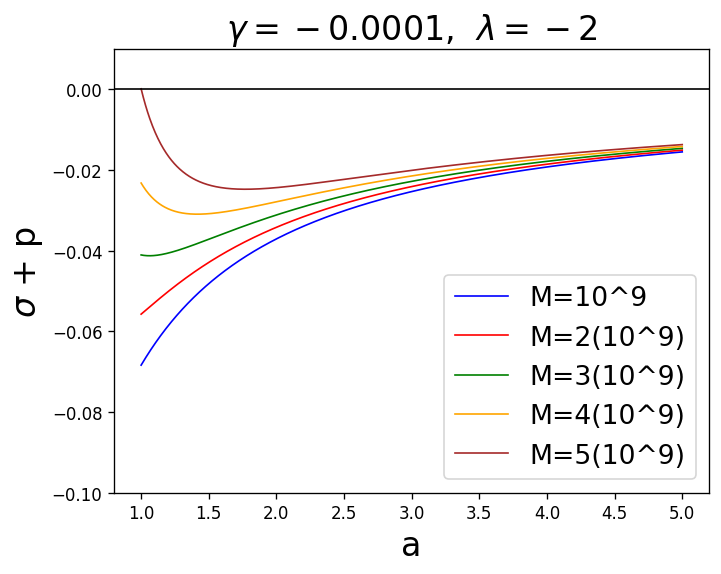}

      \caption{} 
     \end{subfigure}
    \hfill
    \begin{subfigure}[b]{0.32\textwidth}
    
        \includegraphics[width=\textwidth]{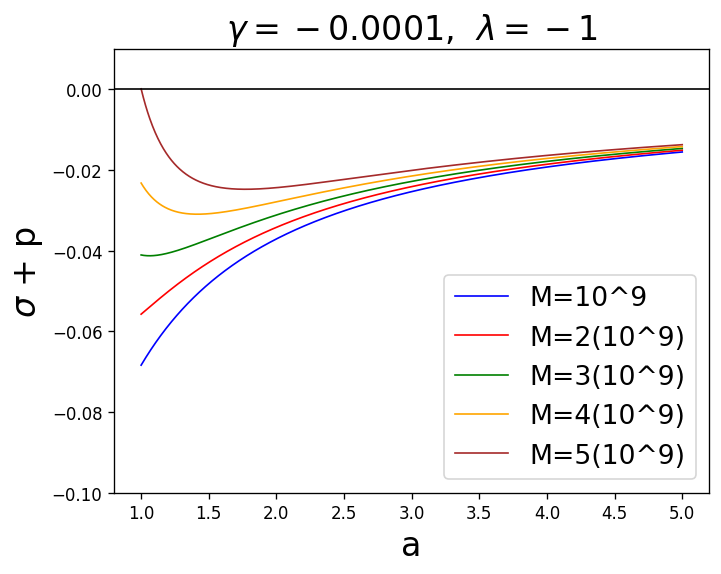}
    \caption{}
    \end{subfigure}
    \hfill
     \begin{subfigure}[b]{0.32\textwidth}
       
        \includegraphics[width=\textwidth]{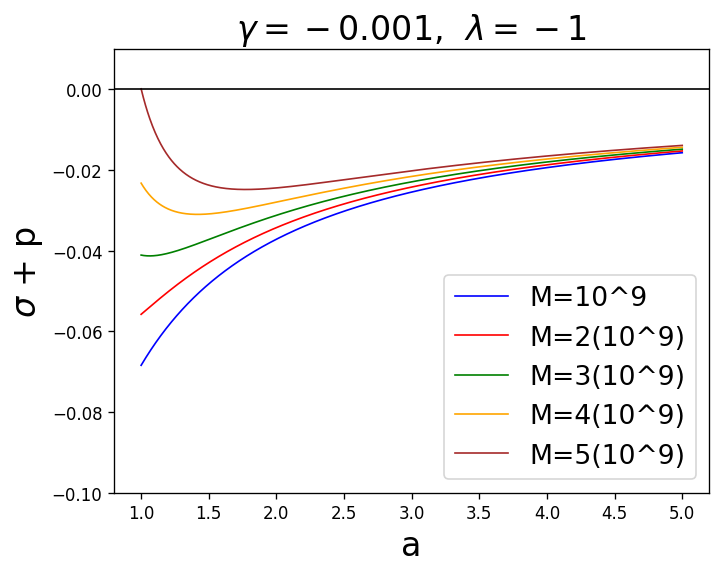}
       \caption{}
    \end{subfigure}
    \caption{The variation of the  \textbf{$\bm{\sigma+p}$} with the throat radius \textbf{$\bm{a(km)}$}. In all three cases, $(\sigma+p) <0$; i.e., NEC is violated. This picture also provides the second condition for the violation of WEC.} 
    \label{fig: sig+p}
\end{figure}

\begin{figure}[H]
\centering
      \begin{subfigure}[b]{0.32\textwidth}
       
        \includegraphics[width=\textwidth]{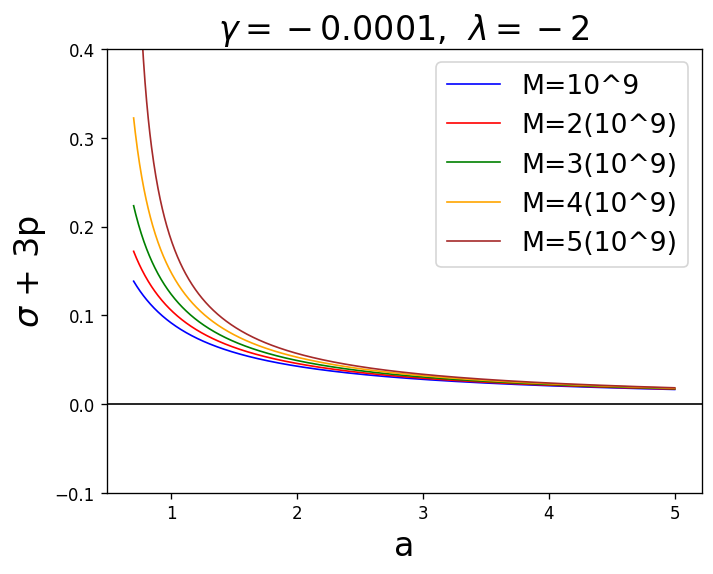}

      \caption{} 
     \end{subfigure}
    \hfill
    \begin{subfigure}[b]{0.32\textwidth}
    
        \includegraphics[width=\textwidth]{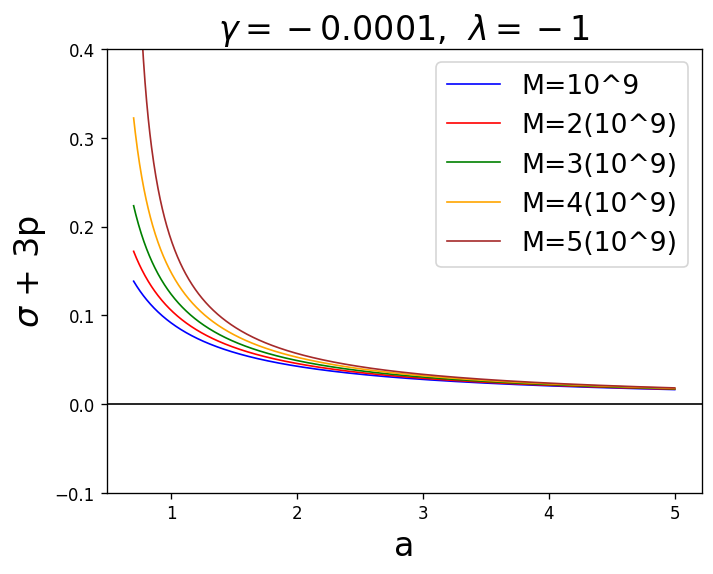}
    \caption{}
    \end{subfigure}
    \hfill
     \begin{subfigure}[b]{0.32\textwidth}
       
        \includegraphics[width=\textwidth]{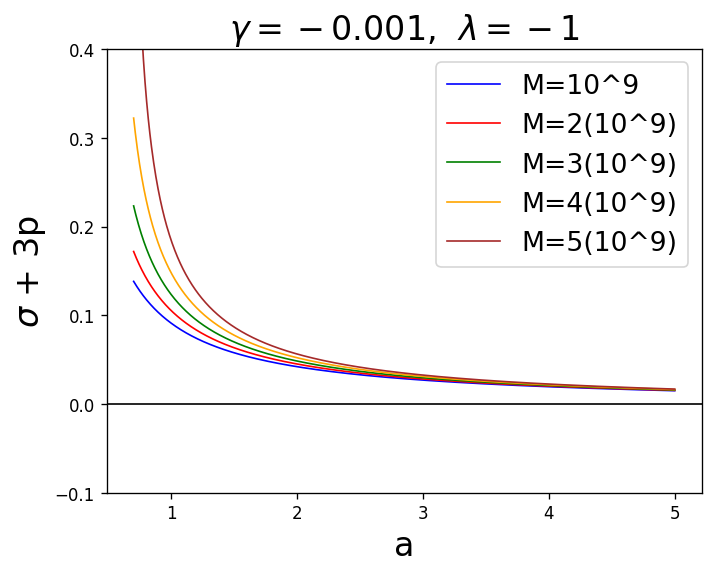}
       \caption{}
    \end{subfigure}
    \caption{The variation of the  \textbf{$\bm{\sigma + 3p}$} with the throat radius \textbf{$\bm{a(km)}$}. The plots show that $(\sigma+3p)>0$; i.e., the wormhole satisfies the Strong Energy Condition.} 
    \label{fig: sig+3p}
\end{figure}

\autoref{fig:sigma} reveals that, with increasing radius, the negative value of the surface stress energy $\sigma$ of a wormhole's throat rapidly decreases at first, and then a very gradual decrease finally leads to almost a saturated value. Additionally, $\sigma$ in the wormholes of smaller masses goes through a sharper change than that of bigger masses, and after a certain radius, its mass dependence vanishes. $p$, on the other hand, (see \autoref{fig: p}) decreases for up to a certain radius, then does not change with mass and radius. Changes in $\gamma$ and $\lambda$ do not affect $\sigma$ and $p$. Violation of the Null Energy Condition (NEC) (i.e. $\sigma + p \ge 0$) at the throat, irrespective of $M$, $\gamma$, and $\lambda$ is shown in \autoref{fig: sig+p}. But the exotic matter in the shell obeys the Strong Energy Condition $\sigma + 3p \ge 0$ (see \autoref{fig: sig+3p} ).


\section{Equation of State (EOS)}\label{sec: omega}
Let the EoS at the surface $\Sigma$ be $p=\omega\sigma$, $\omega \equiv$ constant. From Eqs. \ref{eq: sigma 1} and \ref{eq: p1}, the EOS parameter $\omega$ can be written as

\begin{equation}
    \omega= \frac{p}{\sigma}=-\frac{1}{2} - \frac{1}{2} .\frac{\frac{GM}{a}-\frac{1}{\lambda}.\frac{\gamma}{a^{2/\lambda}}}{1-\frac{2GM}{a} + \frac{\gamma}{a^{2/\lambda}}}.
\end{equation}

\begin{figure}[htbp]
\centering
      \begin{subfigure}[b]{0.32\textwidth}
       
        \includegraphics[width=\textwidth]{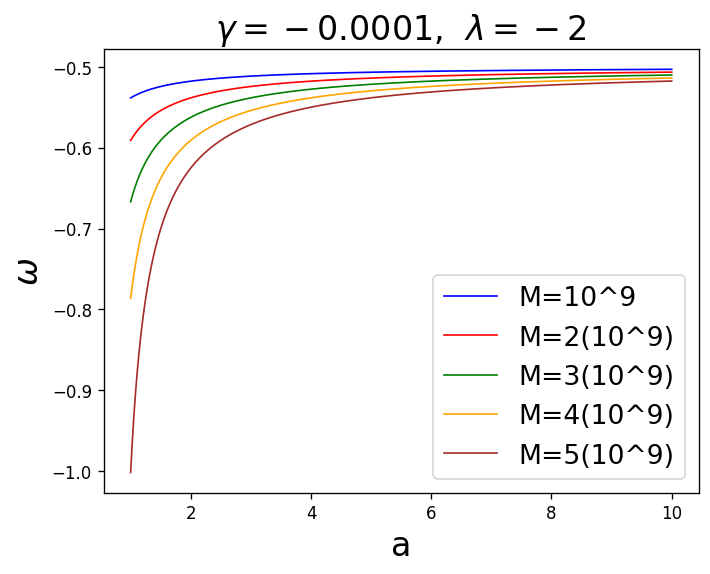}

      \caption{} 
     \end{subfigure}
    \hfill
    \begin{subfigure}[b]{0.32\textwidth}
    
        \includegraphics[width=\textwidth]{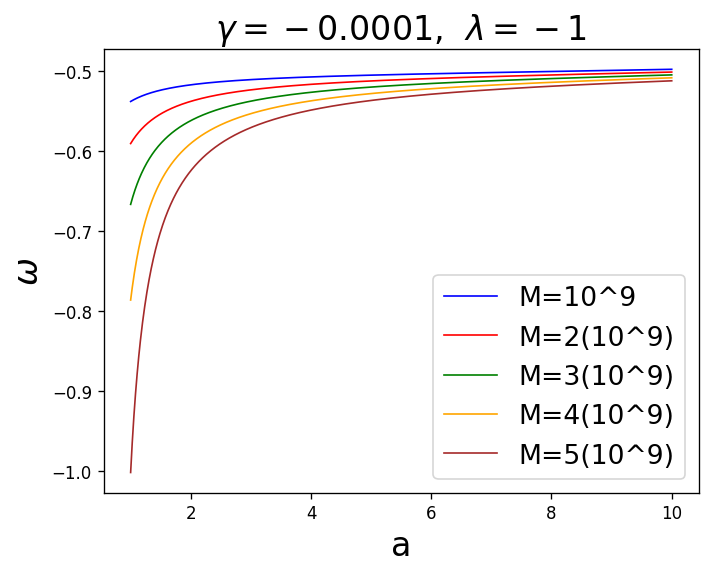}
    \caption{}
    \end{subfigure}
    \hfill
     \begin{subfigure}[b]{0.32\textwidth}
       
        \includegraphics[width=\textwidth]{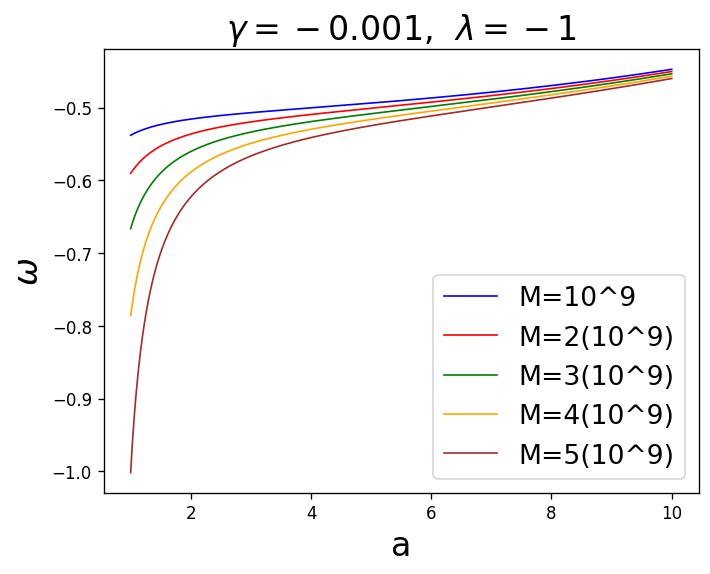}
       \caption{}
    \end{subfigure}
    \caption{The variation of the EOS parameter \textbf{$\bm{\omega=(p/\sigma)}$}  with the throat radius \textbf{$\bm{a(km)}$}. A close observation reveals that $-1<\omega<-\frac{1}{3}$, so the shell is supported by dark-energy-like matter.} 
    \label{fig: omega}
\end{figure}

\autoref{fig: omega} tells us that $\omega$ is negative. Its negative value decreases as the throat radius is increased, and after a point, it is mass-independent. Change in $\lambda$ does not make a significant difference, while change in $\gamma$ does a little ((b) and (c)).

Moreover, it is well known that $\omega>0$ implies the presence of ordinary matter, $-1<\omega<-\frac{1}{3}$ indicates the presence of dark-energy-like matter in the shell, while $\omega<-1$ refers to the phantom-energy-like matter \cite{kuhfittig2016stabilitythinshellwormholesphantomlike}. From \autoref{fig: omega}, it is clear that our thin shell is made of dark-energy-like matter.

Note that if $a$ $\to$ 0 i.e., the wormhole's throat is extremely large, then $\omega \to -\frac{1}{2}$. Additionally, we check for the Casimir effect. Since the Casimir effect with a massless field (KR field in this case) is of
the traceless type, we put $S{^i_j} = 0$ i.e., 
$ 2p - \sigma = 0$, . From this equation, we find that
\begin{equation}
    g(a) \equiv 2- \frac{3GM}{a} + (2- \frac{1}{\lambda})\frac{\gamma}{a^{2/\lambda}} = 0.
\end{equation}

\begin{figure}[htbp]
\centering
      \begin{subfigure}[b]{0.32\textwidth}
       
        \includegraphics[width=\textwidth]{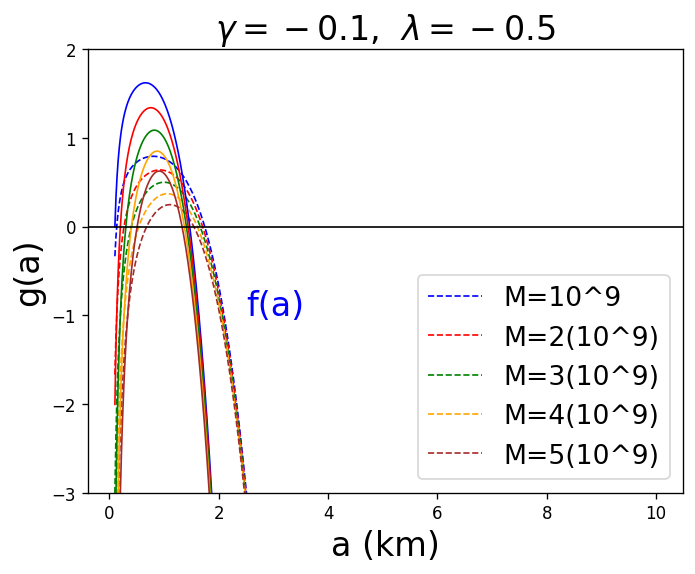}

      \caption{} 
     \end{subfigure}
    \hfill
    \begin{subfigure}[b]{0.32\textwidth}
    
        \includegraphics[width=\textwidth]{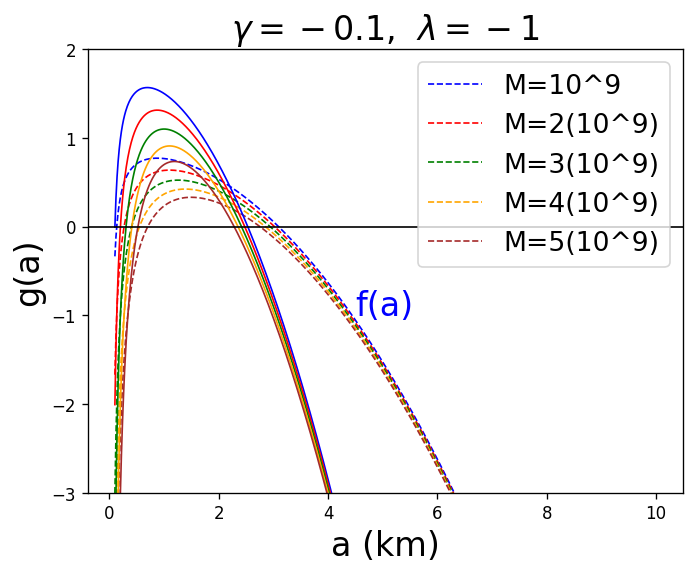}
    \caption{}
    \end{subfigure}
    \hfill
     \begin{subfigure}[b]{0.32\textwidth}
       
        \includegraphics[width=\textwidth]{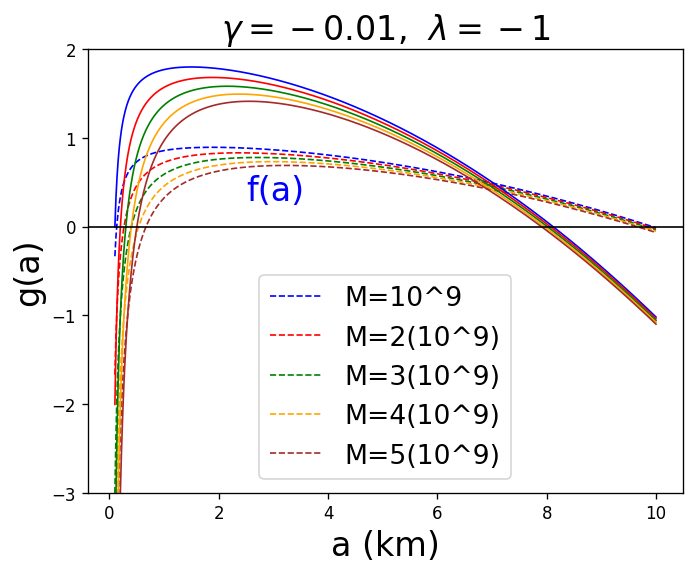}
       \caption{}
    \end{subfigure}
    \caption{The variation of the function \textbf{$\bm{g(a)}$}  with the throat radius \textbf{$\bm{a(km)}$}. For reference, $f(a)$ for each case is also shown with the dashed lines. Clearly, $r_+$ (outer radius) for each $g(a)$ (solid lines) is less than the corresponding $r_+$ for $f(a)$.}
    \label{fig: g}
\end{figure}
Plotting g(a) in \autoref{fig: g}, we observe that the values of $a$ satisfying these equations lie inside the
event horizon $r_h$ (which is actually the $r_+$ corresponding to $f(a)$, as mentioned earlier). However, we have eliminated this portion while constructing our wormhole model. Consequently, this observation rules out the possibility of the Casimir effect for thin-shell wormholes in the KR field.
\section{The Gravitational Field}\label{a^r}
To investigate the nature (attractive or repulsive) of the gravitational field of our wormhole model, we derive the observer’s four acceleration, given by
\begin{equation}
    a^\mu=u^{\mu}_{;\nu} u^\nu,
\end{equation}
where
\begin{equation}
    u^\mu=\frac{dx^\nu}{d\tau}=({\frac{1}{\sqrt{f(r)}},0,0,0}).
\end{equation}

Using Eq. (2), the only nonzero component comes out to be
\begin{equation}
     a^r= \Gamma{^r_{tt}} (\frac{dt}{d\tau})^2 = \frac{f'(r)}{2}=\alpha(r),
\end{equation}
where
\begin{equation}\label{eq: alpha}
    \alpha(r)=\frac{GM}{r^2} - \frac{1}{\lambda} \frac{\gamma}{r^{1 + \frac{2}{\lambda}}}.
\end{equation}

A test particle moving radially from rest satisfies the geodesic equation

\begin{equation}
    \frac{d^2r}{d\tau^2}= -\Gamma{^r_{tt}}{({\frac{dt}{d\tau}})^2}=-a^r.
\end{equation}

\begin{figure}[htbp]
\centering
      \begin{subfigure}[b]{0.32\textwidth}
       
        \includegraphics[width=\textwidth]{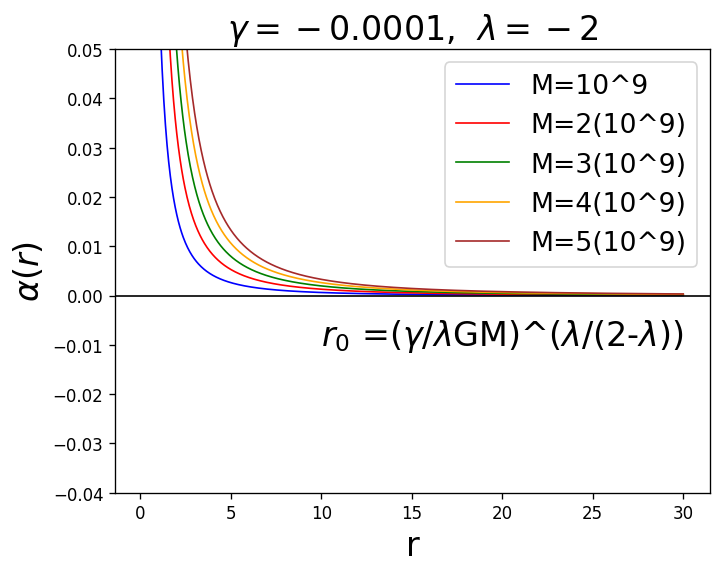}

      \caption{} 
     \end{subfigure}
    \hfill
    \begin{subfigure}[b]{0.32\textwidth}
    
        \includegraphics[width=\textwidth]{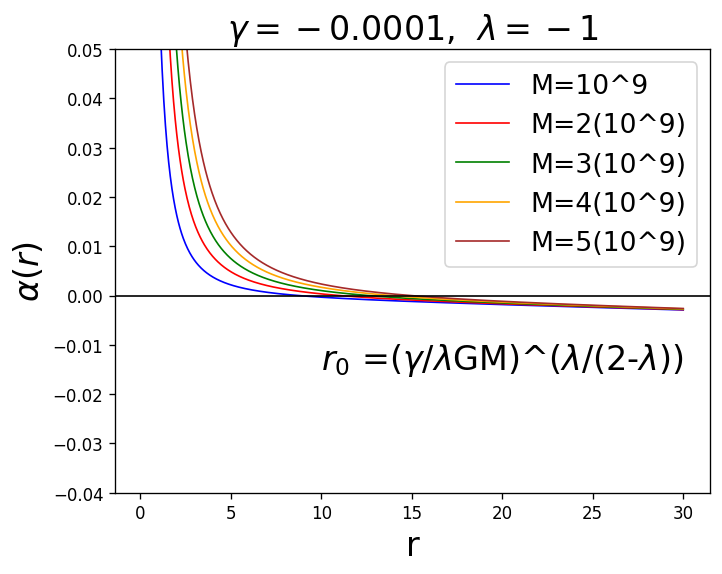}
    \caption{}
    \end{subfigure}
    \hfill
     \begin{subfigure}[b]{0.32\textwidth}
       
        \includegraphics[width=\textwidth]{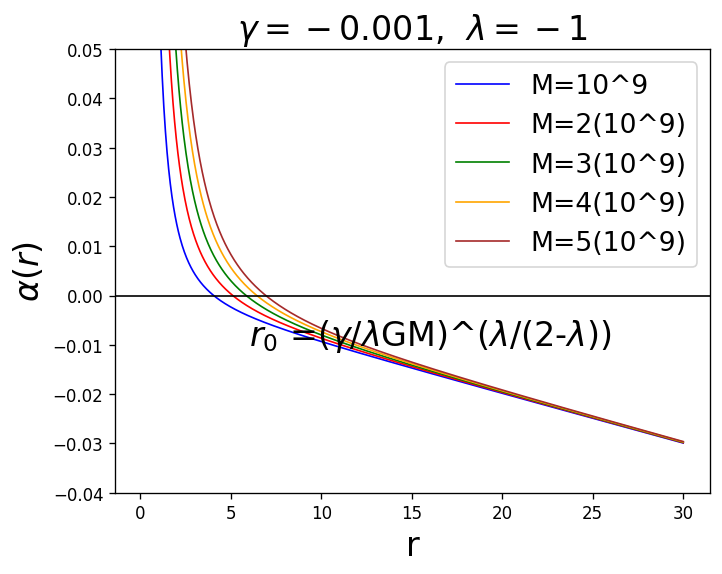}
       \caption{}
    \end{subfigure}
    \caption{The variation of the  \textbf{$\bm{\alpha(r)}$}  with the throat radius \textbf{$r$}, for various masses and LV parameters. $r_0$ is the point where $\alpha(r)$ cuts the $r$ curve. Each curve has a different intersection point; we have pointed out the general expression of the intersecting points.} 
    \label{fig: a}
\end{figure}

In general, a wormhole is attractive for $a^r>0$ and repulsive for $a^r<0$. The Eq. \ref{eq: alpha} and the \autoref{fig: a} shows that $a^r$= $\alpha(r)$ is positive for 
$r< ({\frac{\gamma}{\lambda GM}})^{\frac{\lambda}{2-\lambda}}$, i.e. the wormhole is attractive in this region. In contrast, it is repulsive for $r> ({\frac{\gamma}{\lambda GM}})^{\frac{\lambda}{2-\lambda}}$, since $\alpha(r)$ is negative here. We call the critical radius $r=({\frac{\gamma}{\lambda GM}})^{\frac{\lambda}{2-\lambda}}$ as $r_0$, where $\alpha(r)$ cuts the $r$-axis. The attractive nature decreases with increasing radius up to $r_0$, and from that point the repulsive nature increases. \autoref{fig: a} also reveals that a greater value of $\lambda$ (fig. (b)) or a smaller value of $\gamma$ (fig. (c)) ensures a rising repulsive property. Moreover, a wormhole of larger mass has a bigger critical radius.

\section{Velocity of the radius of the throat}\label{Sec: Rad. Vel}
The Ref. \cite{PhysRevD.70.044008} tells us that the equation of state is independent of $a(\tau)$, i.e., it is the same for a static or a dynamic radius. Using this concept, and following \cite{rahaman2008conical}, we can consider the time evolution of the throat radius and find its velocity.

For static radius i.e. $\dot{a}$ = 0 and $\ddot{a}$ = 0, we know from Eqs. \ref{eq: sigma 1} and \ref{eq: p1} that the equation of state parameter for the static case is,
\begin{equation}
    \omega(a)_s= \frac{v}{\sigma}=\frac{1}{2} + \frac{1}{2} .\frac{\frac{GM}{a}-\frac{1}{\lambda}.\frac{\gamma}{a^{2/\lambda}}}{1-\frac{2GM}{a} + \frac{\gamma}{a^{2/\lambda}}},
\end{equation}
where $v$= the tension at the throat=$-p$.

Now, from Eq. \ref{eq: sigma} and Eq. \ref{eq: p}, the equation of state parameter for a dynamic radius is,
\begin{equation}
    \omega(a)_d=\frac{1}{2} + \frac{1}{2} .\frac{a\ddot{a} + \frac{GM}{a}-\frac{1}{\lambda}.\frac{\gamma}{a^{2/\lambda}}}{1-\frac{2GM}{a} + \frac{\gamma}{a^{2/\lambda}} + \dot{a}^2}.
\end{equation}

So now, equating $\omega(a)_s$ and $\omega(a)_d$, we get the expression for the motion of the throat radius as:

\begin{equation}
   \ddot{a}(a- \frac{2GM}{a} + \frac{\gamma}{a^{2/\lambda}}) = \dot{a}^2(\frac{GM}{a} + \frac{\gamma}{\gamma a^{2/\lambda}}).
\end{equation}
From this equation, we can deduct the velocity (assuming $\lambda=-2$) of the throat radius as:

\begin{equation}
    \dot{a}(\tau) = [\frac{1}{\dot{a}(\tau_0)} - \frac{a(\tau)\gamma}{2(1+\gamma)}-\frac{\sqrt{GM}tanh^{-1}(\frac{a(\tau)\sqrt{(1+\gamma)}}{\sqrt{2GM}})}{(1+\gamma)\sqrt{2(1+\gamma)}} + \frac{\sqrt{GM}tanh^{-1}(\frac{a(\tau_0)\sqrt{(1+\gamma)}}{\sqrt{2GM}})}{(1+\gamma)\sqrt{2(1+\gamma)}}]^{-1},
\end{equation}
where $\tau_0$ is an arbitrary fixed time.

\begin{figure}[h]
    \centering
    \includegraphics[width=0.5\textwidth]{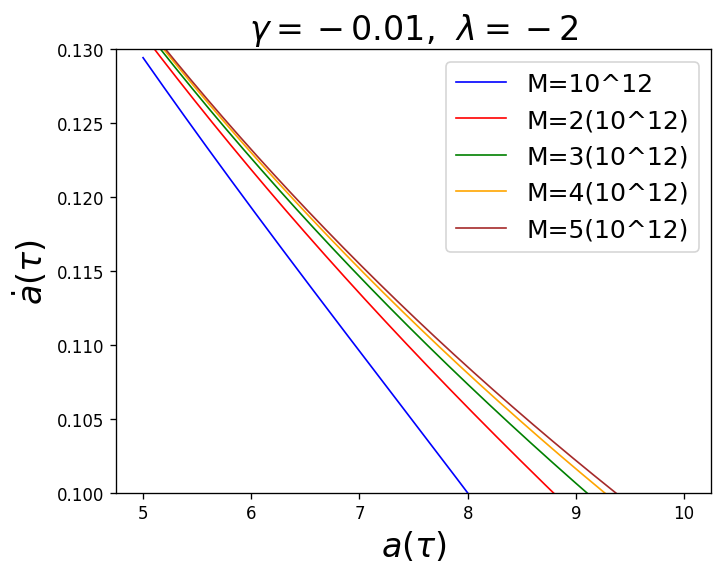}
    \caption{The nature of the velocity of the radius of the throat at time $\tau$ for different throat radii at that moment.}
    \label{Rad. Vel.}
\end{figure}

\autoref{Rad. Vel.} shows that the instantaneous velocity of the throat radius decreases with an increase in the throat radius. Additionally, we observe that the larger the mass of the wormhole, the higher the radius velocity, and the slower the decay of velocity with increasing radius.

\section{The Total Amount of Exotic Matter}\label{Omega}
 The exotic matter count at the junction $\Sigma$ of the thin-shell wormhole can be expressed as \cite{TSW_EMT_GB, TIdalCharged_BH, Heterotic_string_th}:
 \begin{equation}
     \Omega_\sigma= \int [\rho + p]\sqrt{-g}d^3x.
 \end{equation}
Introducing the radial coordinate $R=r-a$,
\begin{equation}
   \Omega_\sigma= \int_{0}^{2\pi}\int_{0}^{\pi} \int_{-\infty}^{\infty}[\rho + p]\sqrt{-g}dR d\theta d\phi.
\end{equation}

Now, since we assume the junction shell is infinitely thin, it does not exert any radial pressure, i.e., p=0, while $\rho$=$\delta(R)\sigma(a)$. Hence,
\begin{equation}
   \Omega_\sigma= \int_{0}^{2\pi}\int_{0}^{\pi}[\rho \sqrt{-g}]|_{r=a} d\theta d\phi = 4\pi a^2 \sigma(a)= -2a \sqrt{1-\frac{2GM}{a} + \frac{\gamma}{a^{2/\lambda}}}.
\end{equation}

\begin{figure}[H]
\centering
      \begin{subfigure}[b]{0.32\textwidth}
       
        \includegraphics[width=\textwidth]{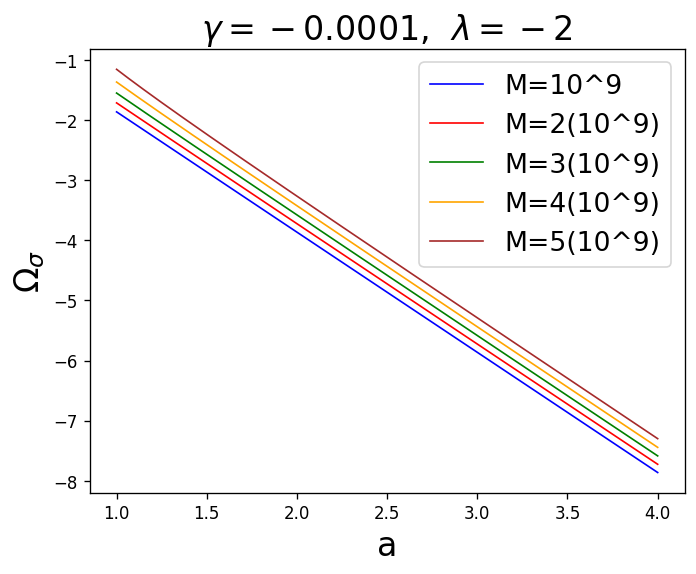}

      \caption{} 
     \end{subfigure}
    \hfill
    \begin{subfigure}[b]{0.32\textwidth}
    
        \includegraphics[width=\textwidth]{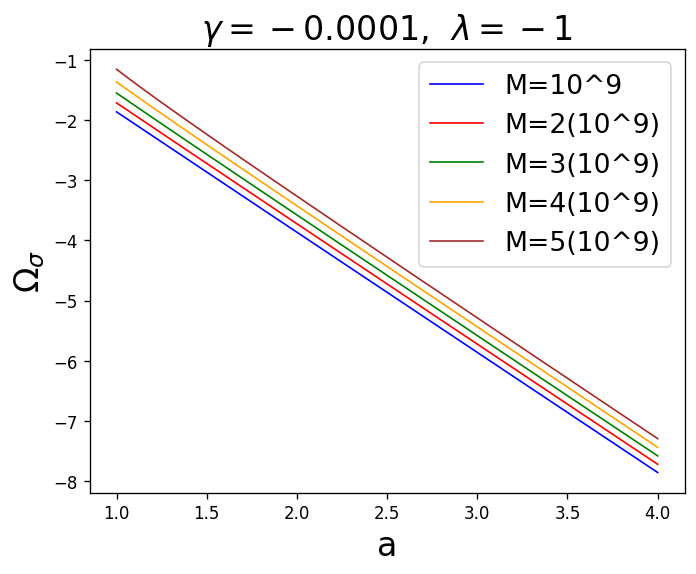}
    \caption{}
    \end{subfigure}
    \hfill
     \begin{subfigure}[b]{0.32\textwidth}
       
        \includegraphics[width=\textwidth]{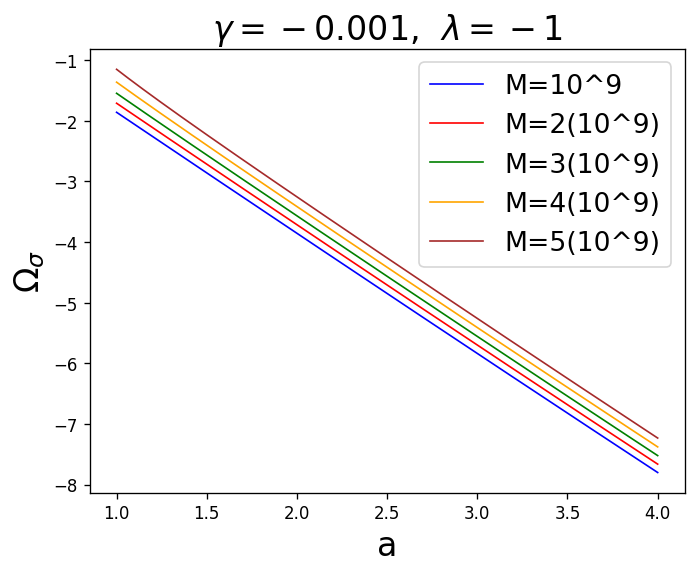}
       \caption{}
    \end{subfigure}
    \caption{The Variation of the total exotic matter at the shell \textbf{$\bm{\Omega_\sigma}$}  with the throat radius \textbf{$a$} (km). } 
    \label{fig: OMEGA}
\end{figure}

\begin{figure}[H]
\centering
      \begin{subfigure}[b]{0.32\textwidth}
       
        \includegraphics[width=\textwidth]{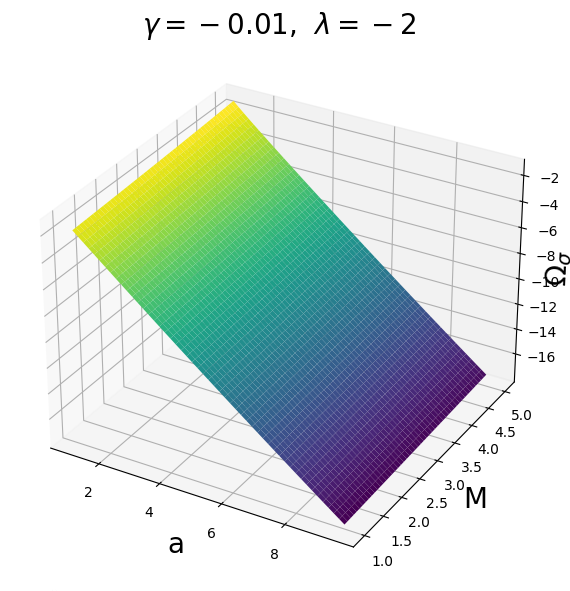}

      \caption{} 
     \end{subfigure}
    \hfill
    \begin{subfigure}[b]{0.32\textwidth}
    
        \includegraphics[width=\textwidth]{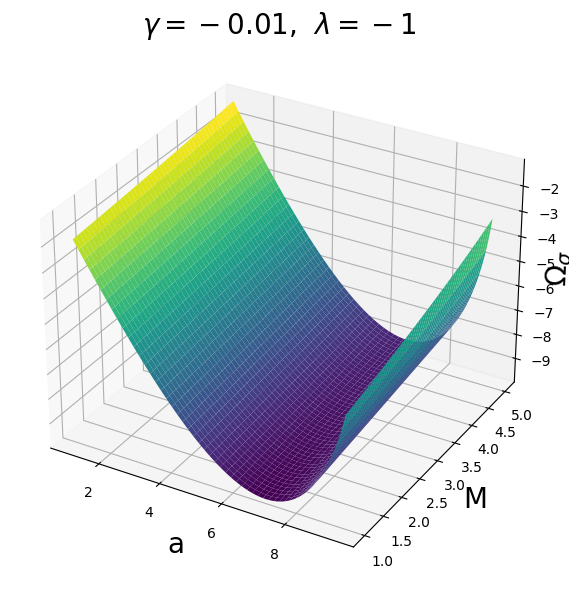}
    \caption{}
    \end{subfigure}
    \hfill
     \begin{subfigure}[b]{0.32\textwidth}
       
        \includegraphics[width=\textwidth]{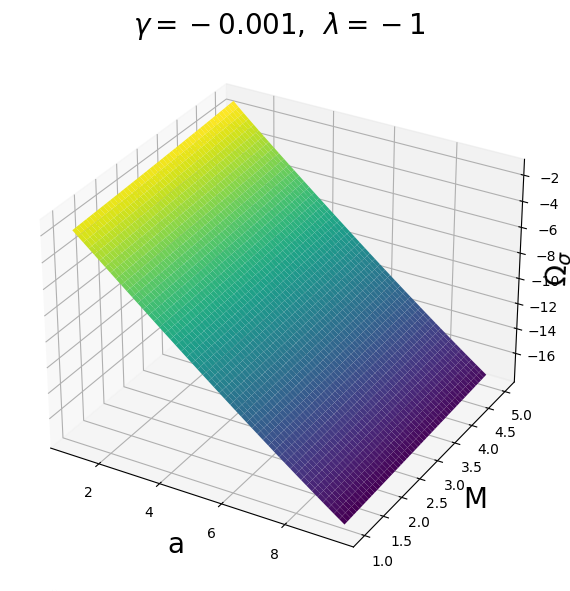}
       \caption{}
    \end{subfigure}
    \caption{The variation of the total exotic matter at the shell \textbf{$\bm{\Omega_\sigma}$}  with respect to the mass of the corresponding black hole \textbf{$M$}, and the throat radius \textbf{$a$} (km). In these 3D plots, the x, y, and z-axis correspond to $a$, $M$, and $\Omega_\sigma$ respectively.} 
    \label{fig: OMEGA_3D}
\end{figure}

\autoref{fig: OMEGA} shows that the exotic matter at the thin shell rapidly decreases with increasing throat radius.  So, in order to minimize it, we should consider the wormhole's throat near the event horizon $r_h$. Also observe that, for $a\gg r_h$, $\Omega_\sigma = -2a$. \autoref{fig: OMEGA_3D} deals with the change in the distribution of the exotic matter for both the mass and the throat radius of the wormhole. We can infer a significant dependence of $\Omega_\sigma$ on the throat radius, while mass-dependence is negligible. Also, small changes in $\Omega_\sigma$ can be noticed due to changes in $\gamma$ and $\lambda$.

\section{Linearized stability}\label{beta}
In this section, we examine the stability of our wormhole setting against a small radial perturbation around a static solution $a=a_0$. First, rearranging the Eq. \ref{eq: sigma}, we obtain the equation of motion at the throat as:
\begin{equation}
      \dot{a_0}^2 + V(a)=0,
\end{equation}
where the potential $V(a)$ can be expressed as
\begin{equation}
    V(a)=f(a) - [2\pi a \sigma(a) ]^2.
\end{equation}\label{eq 26}
Now, expanding $V(a)$ into Taylor series around $a_0$,
\begin{equation}
    V(a)=V(a_0)+V'(a_0)[a-a_0]+\frac{1}{2}V''(a_0)[a-a_0]^2+O[(a-a_0)]^3,
\end{equation}
where the prime ($'$) represents the derivative with respect to $a$. 

Since the linearization is performed about $a=a_0$, we have $V(a)=0$ and $V'(a)=0$. The wormhole attains a stable equilibrium for $V''(a_0)>0$. At this point, we introduce a parameter $\beta$, which we usually interpret as the subluminal speed of sound, and which is given by the relation 

\begin{equation}
    \frac{dp}{d\sigma}\mid_\sigma=\beta^2(\sigma).
\end{equation}
Now, before moving to investigate the conditions for which $V''(a_0)>0$ holds, we derive that $(a\sigma)'=-(\sigma+2p)$ from Eq. \ref{eq 12}. This implies,
\begin{equation}
    (a\sigma)''=-(\sigma'+2p')
=\sigma'( 1+ 2\frac{\partial p}{\partial\sigma})=2(1+2\frac{\partial p}{\partial\sigma})\frac{\sigma+p}{a}=2(1+2\beta^2)\frac{\sigma+p}{a}.
\end{equation}
Going back to Eq. \ref{eq 26}, two successive differentiations w.r.t. $a$ yield,
\begin{equation}
    V'(a)=f'(a)+8\pi^2 a\sigma[\sigma+2p]
\end{equation}
and
\begin{equation}
    V{\rq}{\rq}(a)=f''(a)-8\pi^2[(\sigma+2p)^{2}+2\sigma(1+\beta^{2})(\sigma+p)].
\end{equation}
We can now easily check our required conditions, i.e., $V(a_0)=0$, and $V'(a_0)=0$ from the above equations. The other condition for stability i.e. $V''(a)>0$ gives,
\begin{equation}
2\sigma(\sigma + p)(1+2\beta^2)< \frac{f''(a_0)}{8\pi^2} - (\sigma + 2p)^2.
\end{equation}
Since we have already shown that $\sigma$ and $\sigma + p$ are both negative, the inequality is retained, and we can write
\begin{equation}
    \beta^2 < \frac{\frac{f''(a_0)}{8\pi^2} - (\sigma + 2p)^2- 2\sigma(\sigma + p)}{2[2\sigma(\sigma + p)]}   .
\end{equation}\label{eq 33}
For these inequalities, 
$f(a_0) =  1 - \frac{2GM}{a_0} + \frac{\gamma}{a_0^{2/\lambda}}$,\\ \\
$f'(a_0)= \frac{2GM}{a_0^2} - \frac{2}{\lambda}\frac{\gamma}{a_0^{\frac{2}{\lambda}+1}}$, and \\ \\
$f''(a_0)= -\frac{4GM}{a_0^3} + \frac{2}{\lambda}(\frac{2}{\lambda}+1)\frac{\gamma}{a_0^{\frac{2}{\lambda}+2}}$.

\begin{figure}[htbp]
\centering
      \begin{subfigure}[b]{0.32\textwidth}
       
        \includegraphics[width=\textwidth]{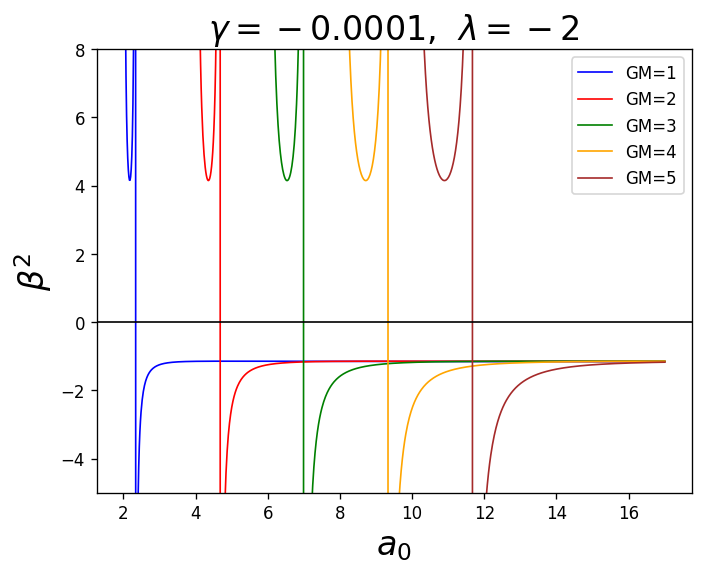}

      \caption{} 
     \end{subfigure}
    \hfill
    \begin{subfigure}[b]{0.32\textwidth}
    
        \includegraphics[width=\textwidth]{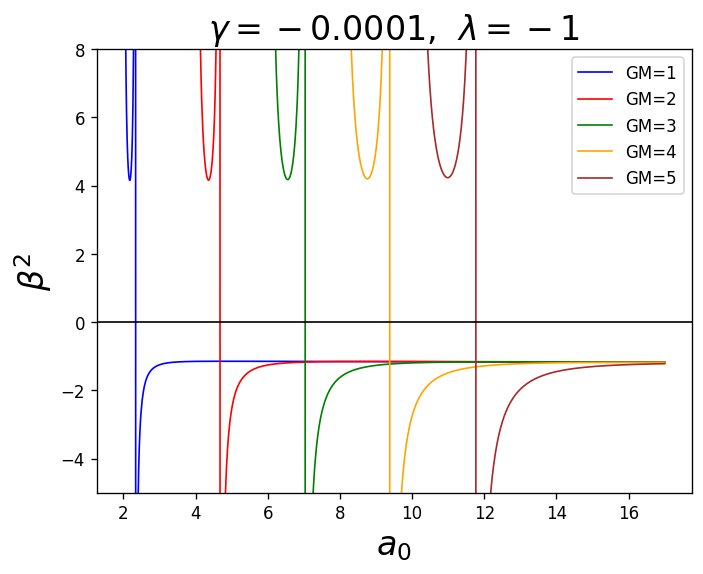}
    \caption{}
    \end{subfigure}
    \hfill
     \begin{subfigure}[b]{0.32\textwidth}
       
        \includegraphics[width=\textwidth]{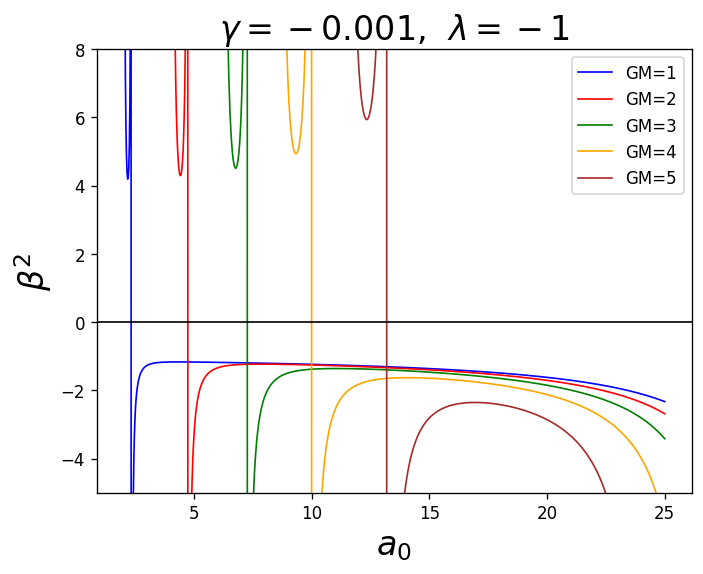}
       \caption{}
    \end{subfigure}
    \caption{The variation of $\bm{\beta^2}$ with \textbf{$\bm{a_0}$}, for various masses and LV parameters. These plots indicate the stability regions of the corresponding wormholes.} 
    \label{fig: beta^2}
\end{figure}

We have plotted $\beta^2$ with respect to $a_0$ in \autoref{fig: beta^2} for five different wormholes with five different masses. According to Eq. \ref{eq 33}, the stability region of each of these wormholes lies under its $\beta^2$ curve indicated in \autoref{fig: beta^2}. We also note that the nature of $\beta^2$ depends on $\gamma$ (fig. (c)).

Now, the fact that $\beta$ is ordinarily interpreted as the speed of sound, imposes a constraint that $0<\beta^2\leq 1$. But \ref{fig: beta^2} shows that the region of stability of our wormhole model involves values out of this range. Hence, we conclude that this wormhole is unstable for all values of $a_0$, when $\beta$ is indeed less than 1. This result coincides with the stability analysis discussed in \cite{10}. However, these conclusions are valid only if $\beta$ is indeed the speed of sound. Moreover, given that exotic matter is needed \cite{10}, and the fact that this shell wormhole scenario is built on a Kalb–Ramond field background, a modified gravitational theory, we should note that the identification made above is uncertain.  Consequently, $\beta^2$ may simply be a convenient fitting parameter without a direct physical interpretation as a sound speed.

\section{Conclusion}\label{Concl.}
This paper investigates a new construction of thin-shell wormholes developed from the non-minimal coupling between the Kalb-Ramond field and the Ricci tensor, thus connecting string theory to the concept of wormholes. This non-minimal coupling gives rise to a power-law-modified black hole solution. We have constructed a thin-shell wormhole by surgically grafting two of these black hole spacetimes using the `cut and paste technique'. Then we explored several salient features of the wormhole both analytically and graphically, while we also looked for the change in the properties with varying LV parameters $\lambda$ and $\gamma$. We observed that some properties significantly depend on one of them, some depend on both, while some remain almost unaffected.

At first, we looked at the nature of the metric function of the given black hole, compared this with the Schwarzschild metric, and thus determined the required values of the  Lorentz-Violating parameters and the minimum radius of the wormhole's throat. This allowed graphical descriptions of energy-density $\sigma$ and pressure $p$ at the thin shell, with respect to the shell-radius $a$ and for different masses of the wormhole, and also for different $\lambda$ and $\gamma$. These results helped us investigate the equation of state, the possibility of the Casimir effect, the attractive or repulsive nature of the wormhole, and the velocity of the throat radius. Then we took a look at the distribution of the exotic matter in the junction surface. Finally, we determined the condition for the wormhole's linearized stability against a small radial perturbation.

An interesting observation is that this wormhole violates the weak and null energy conditions, but obeys the strong energy condition. Another intriguing finding is that the same construction with a radius smaller than a certain value is attractive, while a larger radius implies a repulsive wormhole. Additionally, we observe that the wormhole is unstable for the usual interpretation of $\beta$ as the speed of sound, for which $0<\beta\leq 1$.

However, we should note that $\beta$ might not actually represent the speed of sound, but be just a convenient parameter. We must also keep in mind that both the thin-shell wormhole and the Kalb-Ramond field are theoretical formulations, and the observational or experimental evidence for them is yet to be found.

As a possible extension of this work, light deflection caused by the constructed thin-shell wormhole can be explored using the Gauss-Bonnet theory.

\section{Acknowledgments} Farook Rahaman would like to thank the authorities of the Inter-University  Centre for Astronomy and Astrophysics (IUCAA), Pune, India, for providing research facilities.  He is also thankful to the SERB, DST, and DST FIST programme (SR/FST/MS-II/2021/101(C))  for ﬁnancial support, respectively. Additionally, Arya Dutta is immensely grateful to the IUCCAA and the GLA University, Mathura, India, for selecting him as a participant in the workshop on `Python programming in astronomy, cosmology and astrophysics'.

\section{Data Availability Statement}
No new data was generated or analyzed in support of this research.

\section{Conflicts of interest}
There is no conflict of interest in this article.

%-------------------------------------------
% References
%-------------------------------------------

% Print bibliography

% \printbibliography
\bibliographystyle{plain}
\bibliography{references.bib}
\end{document}